\documentclass[12pt,preprint]{aastex}

\usepackage{emulateapj5}           % emulates original format
\usepackage{natbib}

\slugcomment{Accepted for ApJL on 5th of January 2007, Draft version
  from \today}

\shorttitle{The Protostar in the massive Infrared Dark Cloud IRDC18223-3}
\shortauthors{Beuther \& Steinacker}

\begin{document}

\title{The Protostar in the massive Infrared Dark Cloud IRDC18223-3} 


\author{H.~Beuther$^1$ \& J.~Steinacker$^{1,2}$}
\altaffiltext{1}{Max-Planck-Institute for Astronomy, K\"onigstuhl 17, 69117 Heidelberg, Germany}
\altaffiltext{2}{Astronomisches Recheninstitut am Zentrum f\"ur Astronomie Heidelberg, M\"onchhofstr. 12-14, 69120 Heidelberg, Germany}
\email{beuther@mpia.de, stein@mpia.de}

\begin{abstract}
  At the onset of high-mass star formation, accreting protostars are
  deeply embedded in massive cores made of gas and dust. Their
  spectral energy distribution is still dominated by the cold dust and
  rises steeply from near-to far-infrared wavelengths.  The young
  massive star-forming region IRDC\,18223-3 is a prototypical
  Infrared-Dark-Cloud with a compact mm continuum core that shows no
  protostellar emission below 8\,$\mu$m.  However, based on outflow
  tracers, early star formation activity was previously inferred for
  this region. Here, we present recent Spitzer observations from the
  MIPSGAL survey that identify the central protostellar object for the
  first time at 24 and 70\,$\mu$m.  Combining the mid- to far-infrared
  data with previous mm continuum observations and the upper limits
  below 8\,$\mu$m, one can infer physical properties of the central
  source.  At least two components with constant gas mass $M$ and dust
  temperature $T$ are necessary: one cold component ($\sim$15\,K and
  $\sim$576\,M$_{\odot}$) that contains most of the mass and
  luminosity, and one warmer component ($\geq$51\,K and
  $\geq$0.01\,M$_{\odot}$) to explain the 24\,$\mu$m data. The
  integrated luminosity of $\sim$177\,L$_{\odot}$ can be used to
  constrain additional parameters of the embedded protostar from the
  turbulent core accretion model for massive star formation.  The data
  of IRDC\,18223-3 are consistent with a massive gas core harboring a
  low-mass protostellar seed of still less than half a solar mass with
  high accretion rates of the order $10^{-4}$\,M$_{\odot}$\,yr$^{-1}$.
  In the framework of this model, the embedded protostar is destined
  to become a massive star at the end of its formation processes.
\end{abstract}


\section{Introduction} 

The onset of massive star formation was elusive to observational
research until recently. Very young regions of massive star formation
contain large amounts of cold gas and dust observable from
far-infrared (FIR) to mm wavelengths. A possibly luminous embedded
object has not yet formed or is obscured by the large optical depth of
the surrounding cold dust at near-/mid-infrared wavelengths (NIR/MIR)
The advent of the Infrared Space Observatory (ISO) and the Midcourse
Space Experiment (MSX) allowed to identify large numbers of such
Infrared Dark Clouds (IRDCs)
\citep{egan1998,bacmann2000,carey2000,simon2006}.
%The SPITZER GLIMPSE survey \citep{benjamin2003} will likely identify
%even more similar regions. 
However, the IRDCs are not a well defined class, but these clouds are
expected to harbor various evolutionary stages.  Adopting the
evolutionary sequence outlined in \citet{beuther2006b}, IRDCs should
be capable to contain genuine High-Mass Starless Cores (HMSCs),
High-Mass Cores harboring accreting Low-/Intermediate-Mass Protostars,
and the youngest High-Mass Protostellar Objects (HMPOs). While the
first group is important to study the physical conditions of massive
cores before the onset of star formation, the latter two stages are
essential to understand the early evolution in massive star formation.
Since the evolutionary timescale to form high-mass stars is short (of
the order $10^5$\,yr, e.g., \citealt{mckee2002}), and the central
evolving protostars are deeply embedded within their natal cores,
observational discrimination between these different evolutionary
stages of IRDCs is a challenging task.

Recently, \citet{beuther2005d} observed with the Plateau de Bure
Interferometer (PdBI) a filamentary IRDC containing a compact 3.2\,mm
continuum peak related to a massive core that remains undetected in
the SPITZER GLIMPSE survey up to 8\,$\mu$m (Fig.~\ref{fig1}, left
panel).  Based on the GLIMPSE non-detection one could speculate that
this may be a genuine HMSC, however, there are additional indicators
leading in the opposite direction.  The Spitzer 4.5\,$\mu$m band shows
weak emission right at the edge of the 3.2\,mm continuum core. This
so-called green-fuzzy emission is usually attributed to shock-excited
H$_2$ emission which is especially prominent in this band
\citep{noriega2004}. Such H$_2$ features can be attributed to early
outflow activity. This scenario is supported by single-dish CO and CS
spectra showing non-Gaussian line-wing emission indicative of
molecular outflows.  Furthermore, NH$_3$(1,1) and (2,2) inversion line
observations revealed relatively high gas temperatures of the order
33\,K which would not be expected in the case of a starless core.
Therefore, \citet{beuther2005d} argue that we are witnessing the onset
of massive star formation. Based on the outflow signatures and the
high NH$_3$ temperatures, a central protostar should have formed
already, but it must still be in a very early evolutionary stage to
remain undetected up to 8\,$\mu$m. The detection of at least three
4.5\,$\mu$m outflow features indicates that even a multiple system may
be embedded in this core. At a temperature of 33\,K and a distance of
$\sim$3.7\,kpc (\citealt{sridha,sridharan2005}\footnote{The near
  kinematic distance was chosen because IRDC at $>$12\,kpc are
  unlikely to be observable. The kinematic distance uncertainty is of
  the order 1\,kpc \citep{brand1993}.}), \citet{beuther2005d} calculated
from the 3.2\,mm continuum flux a mass and column density of the gas
core of $\sim$184\,M$_{\odot}$ and $\sim$10$^{24}$\,cm$^{-2}$
($A_v\sim 1000$).

In the framework of the above mentioned evolutionary sequence,
IRDC\,18223-3 should be part of the High-Mass Cores harboring
accreting Low-/Intermediate-Mass Protostars. For such a source, it is
expected that the spectral energy distribution (SED) rises sharply at
mid- to far-infrared wavelength and should hence become detectable
there. The newly released Spitzer legacy survey MIPSGAL (MIPS Inner
Galactic Plane Survey) is the ideal resource to search for such
MIR/FIR emission and study this young source in more detail.

\section{Data}

The MIPS 24 and 70\,$\mu$m data were taken from the Spitzer archive of
the recently released MIPSGAL survey \citep{carey2005}. Fluxes were
extracted via aperture photometry subtracting the background emission
from a close-by region. The Spitzer IRAC 3.5 to 8\,$\mu$m and the PdBI
3.2\,mm continuum data were first presented in \citet{beuther2005d}.
Furthermore, we use a 1.2\,mm continuum flux measurement observed with
the IRAM 30\,m telescope \citep{beuther2002a}. For the 3.2\,mm data,
we use the flux within the 50\% contour level (radius of $\sim
10000$\,AU) to avoid contamination from the large-scale filament.
Similarly, for the 1.2\,mm data we use the peak flux measurement
within the central $11''$ beam. \citet{beuther2005d} estimated for the
whole large-scale filamentary structure that the PdBI observations
suffer from about 25\% of missing flux.  However, since here we are
not interested in the whole filament but only analyze the flux from
the central compact peak-emission, this effect is considerably lower.
The accuracy of the flux measurements at mm wavelength is estimated
from the data to be correct within $\sim$15\% and for the 24 and
70\,$\mu$m within 20\%.  The 3$\sigma$ upper limits of the four
Spitzer IRAC datasets are 0.05\,mJy at 3.6 and 4.5\,$\mu$m, 0.13\,mJy
at 5.8\,$\mu$m and 0.15 at 8\,$\mu$m \citep{beuther2005d}.

\section{Results}

Figure \ref{fig1} presents overlays of the MIPS/IRAC mid-/far-infrared
data and the 3.2\,mm observations. The most striking result is that
the 3.2\,mm core, which is dark at least up to 8\,$\mu$m, is now
detected in the 24 and 70\,$\mu$m bands.  Previously, we could only
indirectly infer from outflow indicators and warm gas temperatures
that the core likely harbors already a very young protostar, but the
new 24 and 70\,$\mu$m data clearly identify this source now for the
first time shortward of 100\,$\mu$m. 

Combining the 3.2 and 1.2\,mm data from the Rayleigh-Jeans part of the
spectrum with the 24 and 70\,$\mu$m fluxes on the Wien-side of the
SED, this dataset allows to derive the physical properties of this
High-Mass Core at the onset of massive star formation in more detail.
Table \ref{fluxes} presents the fluxes from 24\,$\mu$m to 3.2\,mm, and
Figure \ref{fig2} shows the resulting SED.

The spectral energy distribution was fitted with Planck black-body
functions accounting for the wavelength-dependent emissivity of the
dust. The assumed dust composition follows \citet{draine1984}. As a
first order approach, we tried to fit the dataset with a
single-component black-body function. While this may work for the
measurements upwards of 70\,$\mu$m, the 24\,$\mu$m point shows
significant excess emission to a single-component fit, and a second
warmer component has to be added. Figure \ref{fig2} presents a
two-component fit to the data.  While most of the flux, mass and
luminosity stems from the cold gas and dust with an approximate
temperature of 15\,K, we find another warm component with a
temperature around $\geq$51\,K. The gas masses associated with the
cold and warm components are $\sim$576 and $\geq$0.01\,M$_{\odot}$,
respectively. Since on the Wien-side of the SED the emission is not
optically thin anymore, we consider the masses and temperatures of the
warm component as lower limits. The mass difference of the cold
component between the new fit and the older mass estimates from
\citet{beuther2005d} are due to the lower dust temperature we now
derive (15\,K versus 33\,K from the NH$_3$ gas observations) and the
different assumptions about the dust grain size distributions
(following now \citealt{draine1984}, whereas previously we used the
approach by \citealt{hildebrand1983}).  Furthermore, the integrated
luminosity one derives from this two-component fit is
$\sim$177\,L$_{\odot}$ (174 and 3\,L$_{\odot}$ in the cold and warm
component, respectively).

\section{Discussion}

Although, we cannot fit the 24\,$\mu$m flux measurements without the
additional warmer component, it is clear that this component does not
contribute significantly to the mass and luminosity of the region.  In
the regime of low-mass star formation, recently there have been
several detections of very-low-luminosity objects (VeLLOs) within
previously believed starless cores
\citep{young2004,dunham2006,bourke2006}. Some of these objects show
clear outflow signatures (L1014, \citealt{bourke2005}; IRAM04191+1522,
\citealt{dunham2006}), and they are believed to be very young low-mass
protostars or brown dwarfs likely associated with accretion disks.  In
the case of IRDC\,18223-3, the region exhibits outflow signatures as
well \citep{beuther2005d}, hence it is likely that the warm component
in this source is produced by an accreting low-mass protostar-disk
system as well. An important difference between the VeLLOs and
IRDC\,18223-3 is that the luminosity of the warm component in
IRDC\,18223-3~-- although it is nearly negligible compared to the
luminosity of the cold component~-- is about 1-2 orders larger than
those of the VeLLOs. Since we are dealing with a massive gas core that
has on average still very low temperatures and a low total luminosity,
IRDC\,18223-3 is a good candidate of being a High-Mass Core with an
embedded low- to intermediate mass protostar that is destined to
become a massive star at the end of its formation processes.

To constrain the status of the embedded source in more detail, we
compare our data with the analytic and the radiation-hydrodynamic
simulations of the turbulent core model for massive star formation
\citep{mckee2002,mckee2003,krumholz2006b}.  At early evolutionary
stages prior to any hydrogen or even deuterium burning, the luminosity
of the regions is completely dominated by the accretion luminosity
caused by the accretion shocks. The theoretical results obtained by
the analytic approach of \citet{mckee2003} and the simulations of
\citet{krumholz2006b} resemble each other well. Some quantitative
relatively small differences are due to different assumptions in the
initial density distributions, different turbulence damping
assumptions and the origin of the core in a shocked filament in the
simulations which delay all result in higher accretion rates in the
radiative-hydrodynamic simulations. Since the latter likely resemble
real cores better, we use the simulations for our comparison.

Figure 5 in \citet{krumholz2006b} presents the accretion rate and
luminosity of the forming massive protostar throughout its evolution.
A notable difference between the simulations and our observations is
that the simulated luminosity is only that of the primary protostar,
whereas the estimated luminosity of IRDC\,18223-3 includes likely
multiple objects as well as potential contributions from external
heating. However, since there is no detected UCH{\sc ii} region or
strong O-star within several pc from IRDC\,18223-3, external heating
contributions should be negligible. Since massive star formation is
usually proceeding in a clustered mode, multiple embedded protostars
below our current angular resolution limit ($5.8''\times 2.4''$ at
3\,mm wavelength corresponding to approximately 15000\,AU,
\citealt{beuther2005d}) cannot be excluded. Nevertheless, it is likely
that the luminosity of the region will be dominated by the most
massive protostellar object which is also confirmed by the simulations
of \citet{krumholz2006b}. Therefore, it appears reasonable to compare
these simulations with our observational results.

The luminosity obtained for IRDC\,18223-3 is reached in the
simulations by \citet{krumholz2006b} already at extremely early times
when the protostellar mass is still well below half a solar mass. The
corresponding accretion rates are of the order
$10^4$\,M$_{\odot}$\,yr$^{-1}$. Therefore, combining the model
predictions with the observed properties of this region indicates that
we are really dealing with a massive gas core that harbors an
embedded, accreting low-mass protostar with accretion rates that are
high enough that it will eventually form a massive star.

Although the data do not allow us to unambiguously exclude that the
embedded low-mass protostar may remain a low-mass object potentially
never forming a massive star, the fact that the gas core is very
massive, and that the luminosity is already relatively high (a genuine
low-mass protostar has orders of magnitude lower accretion rates and
hence a significantly lower accretion luminosity), strongly supports
the proposed scenario of a massive star-forming region right at the
onset of protostellar evolution and accretion.

The example of IRDC\,18223-3 shows the power of combining (sub)mm
high-spatial-resolution observations with the recently available
near-/mid/and far-infrared surveys of the Galactic plane by Spitzer.
Combining these datasets allows us to constrain the SEDs of the
youngest massive star-forming regions in unprecedented detail. This
way we can observationally characterize the various evolutionary
stages right at the beginning of massive star formation, in particular
the largely unknown properties of genuine high-mass starless cores and
high-mass cores harboring low- to intermediate-mass protostars like
the case of IRDC\,18223-3.  We are currently performing radiative
transfer modeling that incorporates age estimates based on the
emerging MIR flux but this is out of the scope of the current paper.
For the coming years we can expect studies of larger IRDC samples
setting constraints on the source properties in a statistical sense.

\acknowledgments{SPITZER is operated by JPL, Caltech under NASA
  contract 1407. We acknowledge H. Linz, J. Bouwman and S. Quanz for
  help with the SPITZER MIPS data. H.B. acknowledges financial support
  by the Emmy-Noether-Programm of the Deutsche Forschungsgemeinschaft
  (DFG, grant BE2578).  }

%%%%%%%%%% to create the bibliography
%\bibliography{/home/beuther/tex/bibliography}   
%\bibliography{/Users/henrikbeuther/paper/bibliography}   
%\bibliographystyle{aa}    % this does the style, aa.bst necessary

\begin{deluxetable}{lr}
\tablecaption{Fluxes \label{fluxes}}
\tablewidth{0pt}
\tablehead{
\colhead{$\lambda$} & \colhead{$S$}\\
\colhead{$\mu$m} & \colhead{mJy}
}
\startdata
24 & $12.1\pm 2.4$ \\
70 & $989\pm 198$ \\
1200 & $290\pm 44$ \\
3200 & $6.7\pm 1.0$
\enddata
%% Text for table notes should follow after the \enddata but before
%% the \end{deluxetable}. Make sure there is at least one \tablenotemark
%% in the table for each \tablenotetext.
%\tablenotetext{a}{\footnotesize line-blend}
\end{deluxetable}

\begin{figure*}
%\begin{center}
%\includegraphics[width=11.2cm]{IRAC_mips.ps}
%\includegraphics[width=5.5cm]{IRAC_mips_70mu.ps}
%\includegraphics[width=18cm]{f1.ps}\\
%\end{center}
\caption{The color scales show Spitzer images at various wavelength.
  The left panel presents a three-color composite with blue
  3.6\,$\mu$m, green 4.5\,$\mu$m, red 8.0\,$\mu$m (adapted from
  \citealt{beuther2005d}). The inlay zooms into the central core
  region. The middle and right panel show the Spitzer 24 and
  70\,$\mu$m images, respectively. The scaling is chosen individually
  to highlight the most prominent structures.  Contours in each panel
  show the 93\,GHz (3.2\,mm) continuum emission observed with the PdBI
  from 1.08\,mJy\,beam$^{-1}$ (3$\sigma$) in 0.72\,mJy\,beam$^{-1}$
  (2$\sigma$) steps \citep{beuther2005d}. The axes are in R.A
  (J2000.0) and decl. (J2000.0). The circles in each panel present the
  Spitzer beam sizes and the ellipse in the left panel presents the
  PdBI 3.2\,mm continuum synthesized beam. A size-ruler is also shown
  in the left panel.}
\label{fig1}
\end{figure*}

\begin{figure}
\begin{center}
\includegraphics[angle=-90,width=8cm]{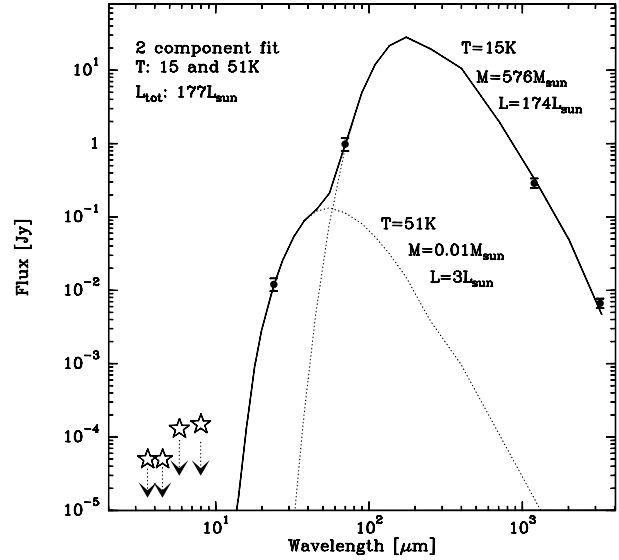}
\end{center}
\caption{Spectral energy distribution of IRDC\,18223-3. The dots with
  error-bars mark the detections at 24 and 70\,$\mu$m on the
  short-wavelength Wien-side of the peak, and at 1.2 and 3.2\,mm on
  the Rayleigh-Jeans part of the spectrum. The four stars below
  10\,$\mu$m show the Spitzer IRAC upper limits in the near-infrared.
  The solid line presents a two-component fit with one cold component
  at $\sim 15$\,K and one warmer component at $\sim 51$\,K. The two
  dotted lines show the two components separately. The resulting
  physical parameters for each component are labeled accordingly.}
\label{fig2}
\end{figure}

\end{document}